\documentclass[journal,12pt, draftclsnofoot, onecolumn]{IEEEtran}

\usepackage[T1]{fontenc}  
\usepackage{url}

\usepackage{color}
\usepackage{subcaption} 

\usepackage{amsmath,amssymb,amsfonts}
\usepackage{algorithmic}
\usepackage{graphicx}
\usepackage{textcomp}
\usepackage{booktabs}
\usepackage{cite}
\usepackage{caption}
\usepackage{mathtools}
\usepackage{multicol}
\usepackage[latin1]{inputenc}
\usepackage{mathtools}
\usepackage{amsmath}

\usepackage{fixltx2e}

\usepackage{array}
\usepackage{booktabs}
\usepackage{multirow}

\begin{document}
\title{Modeling of Viral Aerosol Transmission and Detection}
\author{{ Maryam~Khalid, Osama~Amin, Sajid Ahmed, Basem Shihada  and Mohamed-Slim~Alouini}

\thanks{ M. Khalid is with Electrical and Computer Engineering Department, Rice University, Houston, TX 77005 USA. E-mail : maryam.khalid@rice.edu. 

O. Amin, B. Shihada and M.-S. Alouini are with CEMSE Division,  King Abdullah University of Science and Technology (KAUST),  Thuwal, Makkah Province, Saudi Arabia. E-mail: \{osama.amin, basem.shihada, slim.alouini\}@kaust.edu.sa.

S. Ahmed is with Electrical Engineering Department, Information Technology University, Lahore 54000, Pakistan. E-mail:  sajid.ahmed@itu.edu.pk.}}%

\maketitle
\begin{abstract}
In this paper, we propose studying the disease spread mechanism in the atmosphere as an engineering problem.  Aerosol transmission is the most significant mode among the viral transmission mechanisms that do not include physical contact,  where airflows carry virus-laden droplets over long distances. Throughout this work, we study the transport of these droplets as a molecular communication problem, where one has no control over the transmission source, but a robust receiver can be designed using bio-sensors. To this end, we present a complete system model and derive an end-to-end mathematical model for the transmission channel under certain constraints and boundary conditions. We derive the system response for both continuous sources such as breathing and jet or impulsive sources such as coughing and sneezing. In addition to transmitter and channel, we assumed a receiver architecture composed of air sampler and Silicon Nanowire field-effect transistor. Then, we formulate a detection problem to maximize the likelihood decision rule and minimize the corresponding missed detection probability. Finally, we present several numerical results to observe the impact of parameters that affect the performance and justify the feasibility of the proposed setup in related applications.

\end{abstract}


\IEEEpeerreviewmaketitle

\section{Introduction}

	\IEEEPARstart{M}{olecuar} communication (MC) is an emerging research area that focuses on the communication processes involving biological entities. Unlike conventional wireless communication, where electromagnetic signals are encoded and transmitted to share information, MC uses molecules as signaling sources. Although this phenomenon is a naturally existing communication mechanism in most living beings, it is only recently that it gained attention in the research community. This interest is attributed to the recent advancements in nanotechnology and the advent of nanoscale biosensors or \textit{nanomachines} that have given a boost to research in this field \cite{nakano2012molecular}. The existing nanomachines are constrained in their capabilities due to their small size, limited energy resources, memory and processing capacity. Thus, in order to perform complex operations, multiple nanomachines need to cooperate together and this is where the concept of MC plays an essential role. Since  links between nanomachines can not be established through existing electromagnetic or optical technology, MC provides this connection  that allows them to form a cooperating network of nanomachines \cite{MC_old}. Research in this domain also paves way for development of artificial networks that can imitate biological networks inside or outside the human body. This will not only help in understanding the working mechanism of complex biological systems such as brain, but also help provide cure for several diseases and disorders that occur due to communication links' malfunction inside the body. \cite{malak2014communication}. Thus, it is envisioned that these advancements can play a vital role in biomedical, environmental and manufacturing applications \cite{nakano2012molecular}. Some recently explored biomedical applications include neural network modeling, development of ICT-inspired treatments \cite{malak2014communication} and  intelligent drug delivery \cite{okonkwo2017molecular}.

MC has been studied from the perspective of not only biomedical applications but also from  communications point of view that focuses on the design of efficient receivers, modulation schemes, coding theory and performance analysis, etc. \cite{farsad2016comprehensive}. It must be noted that the existing solutions for conventional communication can not be simply replicated to MC setups due to the complex nature of the process. The challenges in MC appear in the form of non-stationary signal-dependent noise, range limitations that allow nanomachines to communicate over distances not more than few micrometers, large propagation delays, issues concerning chemical reactivity of molecules resulting in high loss rate, limited memory, power constraints and compatibility between bio-nanomachines \cite{nakano2012molecular}. These challenges significantly define the current and future research directions in this area. Furthermore, the fact that these nano-sized sensing bodies allow interaction with biological entities such as bacteria  have also opened several new research avenues. For instance, instead of artificially produced molecules and chemicals, chemicals produced by bacteria serve as messenger for communication between bacterial colonies where receptor bacteria produces light in response to received molecules \cite{bacteria}. In addition to micro and macro-level applications, researchers have also put efforts in understanding and  replicating the existing biological processes/systems and interfacing with them.

In this work, we propose a new dimension in MC that focuses on the spread of infections and diseases via aerosols. Viral aerosols are virus-laden droplets that are suspended in air for prolonged periods of time \cite{def1}. These particles are dispersed in the surrounding because of molecular diffusion and are carried away by wind and this transport is called aerosol transmission. This transmission of viruses leads to disease spread on a very large scale with a massive impact on human population. It has been shown that aerosol transmission is an important mode of transmission for several viruses such as influeza A virus \cite{flu}, severe acute respiratory syndrome (SARS) virus \cite{SARS}, lyssavirus \cite{lyv}, rabies \cite{rab1} and many other pandemics. Unlike the traditional research in MC, for this particular context the message-bearing entities can not be modulated and the message can not be embedded as desired. However, we believe that the virus-laden exhaled air from an infected person can serve as a source of useful information and we need to design our receiver in order to retrieve this information. The significance of this proposed research dimension is even more highlighted in high human population scenarios.

It is common to observe Mass gatherings when people get together for sports, recreational, social or religious activities.  During these gatherings, the large movement of people from different regions poses high risk of disease transmission and transport of emerging and reemerging diseases to the gathering place. The increase in likelihood of disease transmission during Mass gatherings is reported in \cite{MG1, MG2, MG3, MG4}. The detection system proposed in this work can help deal with this problem. If an efficient detection setup is deployed at the entry point of gathering events like railways stations and airports and the likely hosts of diseases and endemics are spotted and treated before they become part of the gathering, the spread of diseases can be significantly prevented. Moreover, if accurate models for virus transport and its dynamics can be established, a blind localization problem can be formulated that can prove helpful in identification of disease sources. Thus, in order to be able take any preventive measures against disease spread, it is essential to characterize and analyze the dynamics of virus transport as has been done in this~work. 
 
We want to highlight that our work is particularly relevant in the context of recent Coronavirus disease breakout (COVID-19). One of the main spread mechanism of this virus is through respiratory droplets released when an infected person coughs or sneezes \cite{world2020management}.  The highly contagious nature of the disease and gaps in our knowledge about its transmissibility, survival duration in the air and spread mechanisms are raising fears about it becoming a global pandemic. This gap can be partly filled by studying the behavior of transmission channel through which the droplets travel. Since person-to-person interaction is the main cause of its spread, there is a need for automated machine-based diagnosis setup so that human involvement is minimized. The scale at which control measures need to be taken also necessitate the deployment of such automated detection infrastructure. This public health emergency requires detection and management of infected individuals in mass gatherings and at points of entry such as international ports, airports and ground crossings \cite{world2020management}. While different corona-detecting biosensors and kits are being developed  \cite{corona3}, concurrent research efforts should be carried out  for their optimal deployment outside laboratory-scale environments.  We believe that the system proposed in this work is a good starting point towards the integration of these biosensors in a  macro-scale virus detection and control infrastructure.

In general, the receiver design proposed in diffusion-based MC is inspired by biosensor technology that aims to detect the presence of a particular biological entity as in \cite{ladhani2017sampling}. However, since MC receivers inherently integrate the biosensors in their functionality, MC receivers are much more extensive and robust in their applications. The biosensors are limited to \textit{detection} of a particular entity only and do not go beyond to extract the information encoded in them in the form of concentration, type, etc. as is the case with MC receivers. MC receivers aim to quantify the information extracted by the biosensor and process them in such a way that information beyond the binary detection can be derived. In addition, there are issues related to the scalability of biosensor-based detection systems especially in macro-scale MC applications such as the one proposed in this work. In general, the research on deployment of biosensors in medicine and virology is limited to laboratory-scale systems. The huge differences between laboratory environment and real-life outdoor environments resulting from physical factors such as presence of wind, infrastructure, etc. discourage the deployment of such biosensors in disease monitoring applications without further exploration and study. Thus, it is very essential that the sensing technology should be complemented with data analysis and processing based on physical domain knowledge. In this work, we not only deal with the aspect of information extraction but also incorporate the effect of scalability to real-life environments by modeling the dynamics of physical aerosol channel.

The well-developed theory in estimation and detection, the presence of molecular receivers, the existence of this phenomenon in nature and the rich set of tools and machinery in wireless communication justifies the treatment of this disease detection problem as a communication problem. Once the modeling process is complete and interface and analogies between this problem and existing wireless communication problem are developed, the efficient set of tools and analysis techniques can be easily deployed to realize systems with capabilities performance unimaginable by biosensor-based approaches.

Being the first to study  viral aerosol information retrieval in communication through breath systems \cite{khalid2019communication}, we investigate the detection problem of a single virus released from an infected human. To this end, we develop a mathematical analysis model of aerosol transmission and detection in order to examine the range limitation of detecting virus.  The main contributions of our work are summarized as follows\footnote{A part of this work was accepted for presentation at IEEE International Conference on Communications \cite{ICC18}.},
\begin{itemize}
	\item A new study dimension in MC is proposed, where virus spread through aerosol transmission is studied as a  MC problem.
	\item A virus detection system is formulated using a biosensor-based receiver architecture and an end-to-end model for proposed setup is presented.
	\item A major part of this work has been dedicated to the modeling of wind-aided aerosol channel that is crucial from the perspective of not only receiver design but also in the study of disease spread. The channel model is derived considering transient analysis, frequency response and steady-state conditions.
	\item The system performance  is studied through numerical simulation scenarios in order to understand the impact of different factors, where the performance is evaluated by studying the probability of virus miss-detection.
	
\end{itemize}

The rest of the paper is organized in the following order: Section \ref{sec:desc} provides a brief description of proposed setup. System modeling is explained in detail and detection problem is formulated in Section \ref{sec:sys}. Modeling of aerosol transmission channel is provided in Section  \ref{channel} followed by simulation and results in Section \ref{result} finally concluding the paper in Section \ref{conc}.

\textit{\textbf{Notations}}: As for the mathematical notations, we use the following symbols. The partial derivative of the function $f$ with respect to $x$, i.e, $\frac{\partial f}{\partial x}$, is represented by 
$f_x$. Higher order derivatives  are represented by repetition of independent variable in subscript, for example $\frac{\partial^2 f}{\partial x \partial y}$ is represented by  $f_{x,y}$.  The $x$, $y$ and $z$ components of $\vec{F}$ are defined as $F^{x}$, $F^{y}$ and $F^{z}$, respectively.  The Laplace transform of a function $f(x,y)$ with respect to $x$ is represented by $\bar{f}^{(x)}(q,y)$. The two dimensional  Laplace transform is represented by  $\bar{\bar{f}}^{(x,y)}(q,r)$. $\vec{F}$ denotes that $F$ is a vector and $\nabla$ is the vector differential operator. Thus, the divergence of $\vec{F}$ is expressed as $\nabla. \vec{F}$.

\section{System Description}
\label{sec:desc}
In this section, we briefly describe the basic architecture of a single source viral aerosol transmission system. The proposed system is composed of three major components. The first one is the infected human who acts as a source of pathogen and in the rest of the paper, he or she is referred to as the transmitter. The second component is the aerosol transmission channel through which  the virus travels. The transmission can be subjected to air-flow, i.e.,  artificial  wind. The third component is the receiver side that aims to retrieve information about the virus and/or pathogen.

The respiratory tract of the infected person is loaded with virus-containing droplets. The virus is emitted from the mouth through different  mechanisms that include breathing, coughing and sneezing.  If we consider the instantaneous emission, coughing and sneezing release higher droplets than breathing. However,  as a process they are not very frequent compared to the normal breathing which is a continuous process \cite{breath}. Thus, over the coarse of long time windows like a few hours or so, coughing and sneezing account for smaller proportion of bioaerosols compared to normal breathing  \cite{breath}. 

\begin{figure}
\centering
 \includegraphics[width= 3.5in]{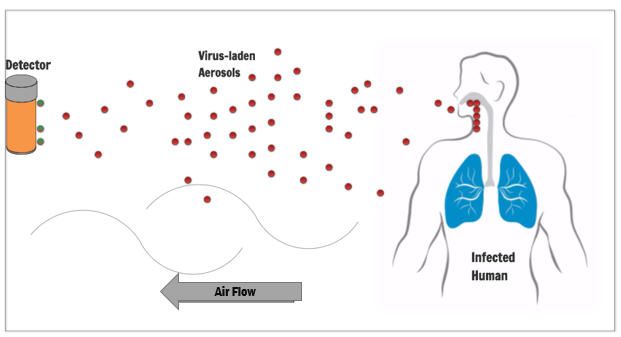}
\caption{Virus aerosol transmission system setup.}
\label{system_setup}
\end{figure}

Once the aerosols are released into the air, they disperse in different directions because of diffusion. To increase the transmission range, decrease the propagation delay and improve the strength of received signal, we apply air flow that is directed towards the receiver side. It must be noted that in real-life situations wind plays an important role in the spread of pathogens and diseases and therefore, incorporating air flow brings our setup close to real-life systems. Along with some other physical properties,  the size of droplets determines the travel distance of the pathogen. Aerosol transmission can be categorized into two types, airborne transmission and droplet transmission. Airborne transmission is defined as transmission of aerosols with size less than $5\mu \rm{m}$, which can travel for a large distance in order of meters \cite{air, air2}. On the other hand, droplet transmission refers to transmission of pathogen-laden droplet whose size is usually greater than $5\mu \rm{m}$ with shorter distance spreading \cite{air2}. Our focus in this paper is airborne transmission.

Throughout this paper, we are interested in retrieving viral information from aerosols released from the respiratory tract of infected humans which is depicted in Fig. \ref{system_setup}. The experiment is performed in an indoor environment where we can apply artificial airflow with a specific velocity to drive the aerosols towards the detector. It must be noted that the experimental setup under consideration is very close to real-life situation where wind is responsible for spread of viral droplets. Thus, the models derived in this work can be deployed not only in bio-monitoring applications but can also prove helpful in the qualitative and quantitative analysis of infection transmission.

\section{System Modeling}\label{sec:sys}
\begin{figure*}[t!]
\centering
\includegraphics[width=5in]{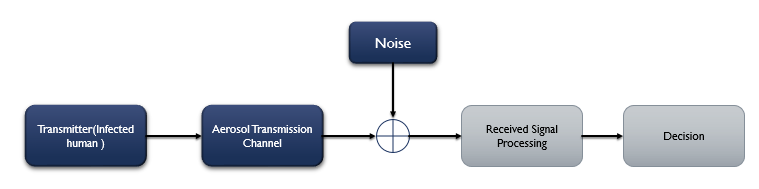}
\caption{System Block Diagram}
\label{block_diagram}
\end{figure*}

The objective of this section is to analyze all blocks of system, shown in Fig. \ref{block_diagram}, separately. The system is composed of a transmitter,  followed by physical channel with additive noise and finally the detector. In the following subsections, we provide the detailed  mathematical modeling of different system components. 

\subsection{Transmitter }\label{input}
  
As explained earlier, we assume that the infected human releases pathogens into the air through his breath. The normal breathing of an adult occurs at rate of 12-16 breaths per minute \cite{rate}, which means that each breath takes no more than 4.98 seconds. Please note that compared to wireless communication, the transmission achieved through chemical signaling is quiet slow and it can take several minutes for the signal to reach the receiver placed few meters away. Thus, the transmission process ensures that the time scale of this experiment is of the order of the minutes.   Since the time within breaths (or exhalation time specifically) is too small
 
\textit{relative} to the coarse of experiment which is of the order of several minutes,  the variations in emission process due to exhalation can be averaged out and the process can be approximated as a continuous and constant emission process. The emission rate might vary over time however we expect the average rate to be constant atleast for the coarse of experiment  (of the order of few minutes). Although it is helpful if the variations in the emission rate can be incorporated in the system design, it is hard to find any such deterministic or stochastic model that can explain these variations in the literature. Most empirical studies that revolve around this subject are based on collection of breath samples of length varying from few minutes ($\sim$ 30 minutes) up to hours and reporting the cumulative effect. It is not clear whether the variations in per-second emission rate for the few minutes time windows can be evaluated with the current technology.  Hence, we  model the input signal as a continuous process with constant average emission rate which is equal to the  average rate i.e $Q$ g/sec.

It is important to note that although breathing can be assumed to be continuous, coughing and sneezing as impulsive jet sources can not follow the same assumption.  
For modeling the input signal, the duration of experiment or temporal characteristics of the application under consideration are extremely important. For some applications such as understanding the dynamics of diseases spread or detection in specific frameworks where only one person enters the room and stays for long enough, the steady state response of the system is sufficient. However, for applications where decisions are made based on data collected at fine-grained resolution, the transient response is required. Since, the latter yields the system impulse response, we start with transient analysis for jet sources in the next section extending it to the transient analysis for breathing and finally moving to the steady state response for breathing. For the rest of the system blocks that include the noise and detection block, we focus on the steady state response only.
For transient analysis, the input is modeled differently to incorporate the dynamics of both, time and space. At this point, we do not comment on the frequency of jet sources, their probability to occur, time interval between successive sneezes and the duration of continuous coughing. We consider a sneeze or single cough as a jet  \textit{instantaneous} source that emits $R_s$ aerosols into the air. If a person of height approximately $H$ standing at location $[0,0,H]$ in Cartesian coordinate space, sneezes/coughs at time $t=0$, the source is modeled as,
\begin{equation}\label{sneeze}
{S_{\mathrm{s}}} = {R_{\mathrm{s}}}   \delta(x)\delta(y)\delta(z-H)\delta(t).
\end{equation}
Similarly, while the person is breathing, he is continuously emitting aerosols with a certain flow rate $R_b$. If the person  has entered the room or experimental setup at time $t=0$ and stands again at the same location $[0,0,H]$, the source is modeled as,
\begin{equation}
{S_{\mathrm{b}}}   = {R_{\mathrm{b}}} \delta(x)\delta(y)\delta(z-H)u(t).
\end{equation}
Since a person who sneezes is also breathing, both continuous and jet sources should be included in the definition of input signal.  
Thus, we assume both these emissions to be independent of each other and define the final input signal to be,
\begin{equation} 
{S_{\mathrm{t}}}= {S_{\mathrm{s}}}+{S_{\mathrm{b}}}.
\end{equation}
If there are multiple people present in the room at different locations, their independence from each other results in simple addition of all those emissions to represent the final input signal.
Along with the emission rate,  the aerosol droplets' size  also affects the communication performance. According to the study in \cite{size}, it has been observed that the normal human breathing results in larger fraction of droplets compared to coughing and sneezing, and the droplet sizes are below $1 \mu \rm{m}$. These micro-sized droplets also have a high tendency to remain suspended in the atmosphere for a long period. Thus, this work circumscribes aerosols that travel distances of the order of meters and remain suspended in air for more than several minutes which is the coarse of our simulation experiment. 

In this work, our major focus is on the breathing only as it is a permanent source of aerosol transmission compared to coughing and sneezing. Moreover, it is not practical to wait for the infected person to cough or sneeze to begin the experiment of detecting and tracking the infection. However, it is still worthwhile to analyze the performance of coughing and sneezing mechanisms and therefore we derive the transient analysis of aerosol channel as well.

\subsection{Molecular Receiver}

We propose a detection mechanism that could be leveraged to distinguish between healthy and infected person. Once, the infected person has released certain amount of pathogens into the atmosphere, and they have passed through the molecular channel, the receiver acts as an absorbing surface that absorbs most of the pathogen -laden droplets. The architecture of molecular receiver is shown in Fig. \ref{fig:arch}. The details of three major blocks are presented below.

\begin{figure}[h!]
	\centering
	\includegraphics[width= 5in]{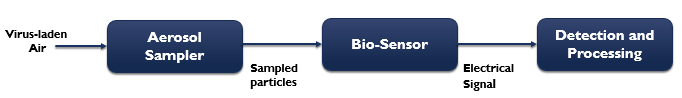}
	\caption{Receiver Architecture}
	\label{fig:arch}
\end{figure} 

\subsubsection{Aerosol Sampler}
Several techniques have been developed for collection of suspended air particles. Aerosol sampler is the front end of our receiver which controls the sampling rate of air. Although there are several other techniques, the sampler proposed in this receiver architecture is based on the principle of electrostatic precipitation which is not only commercially available but also allows  sampling of particles with sizes as small as 2-100nm.\cite{sampler}. The sampler sensitivity in terms of sampling nano-sized particles is quiet significant since, the droplet sizes are of the order of few micrometers and the diameter of virus and bacteria, in general can be of the order of nanometers. 
\begin{figure}
	\centering
	\includegraphics[width=3.5in]{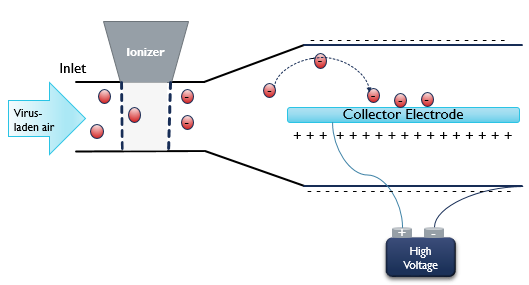}
	\caption{Electrostatic air sampler.}
	\label{sampler}
\end{figure} 

The architecture of the electrostatic air sampler is depicted in Fig. \ref{sampler}. The two main components of the sampler are ionizer and charged electrode. The ionizer induces negative charge on air particles that pass then through the next chamber and collect on the positively charged electrode after repelling by the outer negatively charged boundaries..
The performance of sampler is quantified through it's collection efficiency. As reported in \cite{sampler2}, collection efficiency of $80-90\%$ is achievable with commercial electrostatic aerosol samplers. For the rest of paper, we denote sampler efficiency by $\xi$.

\subsubsection{Biosensor}
Biosensors are sensing devices that translate molecular  events into processable information \cite{bios, bios2}. They usually consist of a recognition layer followed by transducer that converts the recognized signal into a processable form \cite{MCr}.
The physical change that the bio-recognition layer undergoes on contact with target bodies is quantified. The recognition layer is connected to a transducer which converts the behavior or the variation in recognition layer into processable information. Based on the varying property which is being measured, biosensors are classified into three types, electrical, mechanical and optical.

In the electrical biosensors, a change in current, voltage, or conductance is observed when a binding event takes place in the recognition layer. On the other hand, in optical sensors, the optical properties of the recognition layer are altered when it comes in the presence of target cells. As for the mechanical biosensors, they consist of nanomechanical systems that are capable of detecting the forces, motions, mechanical properties and masses, which  emerge in biomolecular interactions \cite{mech}.

In this work, we consider electrical biosensors due to their high sensitivity and selectivity \cite{bios_e}. There are two basic types of electrical biosensors, which are known as biocatalyitic and affinity based. In biocatalytic sensors, the presence of a virus or a target specie induces enzymes (already present on recognition layer) to produce a certain chemical substance whose concentration is then measured in order to obtain information about the presence of target. On the other hand, affinity based sensors consist of virus or target-specific antibodies placed on the recognition element. In the presence of target, a binding event between target and antibody takes place which is translated to variation in some electrical property (current, voltage, conductance, etc.). Field effect transistors (FETs) are the most commonly used affinity-based electrical biosensors. The amplification property of FETs permits that a small change in voltage at the gate nduces a large current change in source-drain channel yielding a highly sensitive biosensor \cite{fet2}. 

\begin{figure}[t]
	\centering
	\includegraphics[width=3.5in]{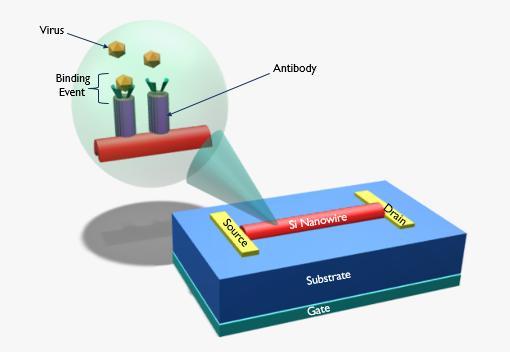}
	\caption{Si-NW FET}
	\label{fet}
\end{figure}

The Silicon NanoWire (Si-NW) FET is transistor with Si-NW is placed between the source and the drain terminals over the FET substrate. The virus is detected with the help of antibody receptors that  are placed on the Si-NW as shown on 
Fig. \ref{fet}. When the FET is placed in an antigen-rich solution, the antigens come into contact with the antibody receptors and a binding event takes place. The binding events effect the source-drain conductance channel resulting in accumulation or depletion of electrons just like gate voltage. Thus, the binding events are translated to current change across the source-drain channels by producing a change in the FET conductance. The inherent amplification property of FETs allows for even small number of binding events to produce measurable current making it a highly sensitive sensor \cite{fet5, fet3, fet4}. It was  shown in \cite{fet5}, that the presence of a virus resulted in a dramatic change in source-drain current. In \cite{fet3}, the presence of influenza A virus results in discrete changes in conductance while no change in conductance occurs in presence of other viruses such as adenovirus, which demonstrates the high selectivity of the sensor. In \cite{fet4}, an air sampler was integrated with FET and it was shown that discrete conductance changes were observed when sensor was exposed to aerosols and the change was proportional to the aerosol concentration. The setup was able to detect in real-time and the whole process of sampling, sensing and detection occurred in less than 2 minutes.

 The capabilities of bio-FETs also depend on material used to fabricate the conductance channel. The two materials that have gained most attention in recent years are silicon-based (Si-NW) and nano-carbon materials such as Graphene and Carbon Nanotubes (CNT). While, SiNW-FETs have shown high sensitivity and selectivity for real-time detection \cite{sinw1} and are well studied in literature, they suffer from low-carrier mobility \cite{sinw3} and deterioration from oxide-layer built-up \cite{sinw2}.  Compared to that, Graphene and CNT offer much better mechanical strength, chemical stability, electrical properties and  plasticity \cite{sinw4}.  However, the latter suffers from reproducibility issues and the synthesis process produces impurities that degrade electrical properties of these materials. The fabrication of SiNW is much easier and unlike nano-carbon materials it can be produced on mass-scale by the semiconductor industry. The mater fabrication process, feasibility in mass-production and the availability of extensive literature  encourages us to focus on SiNW-FETs in our work.

\subsubsection{Measurement Noise} 
The most important factor while designing the processing and detection block is the input signal model and noise. As explained in \cite{noise2}, binding noise and flicker noise are dominant in MC receivers.
\begin{itemize}
	
	\item \textit{ Binding noise}: The probabilistic binding events between virus and receptors existing at the FET biosensor result in a binding noise. A detailed modeling of the binding noise is discussed in \cite{noise3, noise4}. A Markovian approach isadopted to model the dynamics of biosensor and derive closed form expressions for settling time and noise power spectral density (PSD). Since, we are interested in analyzing the system in the steady state, where we  wait for a longer period than the settling time before measuring the output signal, the captured virus concentration   is given by \cite{noise4},
	\begin{equation}
	C_{\rm{ss}} = N \frac{P_{\rm{a}}}{P_{\rm{a}} + K P_{\rm{d}} } = N \gamma 
	\end{equation}
	where $N$ is the total number of antigens, $P_{\rm{a}}$, $P_{\rm{d}}$, and $K$ are the association probability, dissociation probability and number of possible states, respectively.
	
	\item \textit{Flicker Noise}: This noise is known also as $1/f$ noise, where it is dominant at low frequencies and results from semiconductor channel imperfections. Flicker noise can be modeled as Gaussian \cite{1f, 1f2}  and is independent of the virus concentration because it solely depends on the transistor characteristics.
	
\end{itemize}

Moreover, we can have interference noise that results from other biological entities which might interfere with the binding process of virus and the corresponding antibody receptor \cite{noise2}.   The accumulating nature of the aerosol channel can also result in significant inter-symbol interference (ISI). A passive solution to overcome the effects of ISI is to introduce large delays between the time a person leaves the room and the next person enters so that the aerosols are dispersed. Another solution can be to incorporate the accumulation effect in the detection mechanism and update the decision threshold each time a person leaves the room. However, a more efficient method is using enzymes that react with accumulated viral aerosols and keep them from interfering in future\cite{Interference}. We assume that such method has been employed to take care of interference effects such that each new person entering the room is exposed to a sanitized environment; thus the interference can be ignored.

\subsection{Aerosol Transmission Channel }
Once the bioaerosols are released into the air by the infected human, the droplets are carried away by the artificial wind, which is applied to increase the transmission range and decrease the propagation delay.  The basic difference between diffusion-based MC (DMC) and aerosol transmission is the propagation mechanism that drives the droplets in the fluid. In DMC, the Brownian motion  is responsible for movement of particles and can be modeled as a \textit{Wiener} process \cite{nakano2012molecular}, where the molecular diffusion is characterized by molecular diffusivity coefficient. This communication occurs due to thermal movements of molecules and therefore it is a micro-scale communication. On the other hand, aerosol communication is a macro-scale transport of micro-sized particles over larger distances and can be characterized by dispersion models. 
Moreover, molecular diffusion has negligible contribution in the propagation of bioaerosols in atmosphere and is mainly governed by advection and turbulent diffusion. The wind is responsible for advection and  eddies cause turbulent diffusion. It must be noted that eddy diffusivity coefficient is much greater than the molecular diffusivity coefficient, which can be ignored in dispersion models. 

Now we describe the basic framework for deriving the channel model. Assume the source is located at $\vec{r}= [x,y,z]$ emitting pathogens at a rate  $S(\vec{r},t)$, where $t$ is the time. From law of conservation of mass we can write \cite{cdiff1, cdiff2},  
\begin{equation} \label{Law_conv_mass}
C_t + \nabla .\vec{F} = S,  
\end{equation}
where $C_t$ is change in concentration of pathogen with time and $\vec{F}$ is the mass flux.  Both, the concentration and flux are functions of $\vec{r}$ and $t$. The mass flux is composed of two components resulting from the two phenomenons called diffusion and advection and can be represented as follows,
\begin{equation}
\vec{F } = \vec{F}_{\rm{diff}}+ \vec{F}_{\rm{adv}},      \nonumber 
\end{equation}
where $\vec{F}_{\rm{diff}}$ is the diffusion component and $\vec{F}_{\rm{adv}}$ is the advection component. 
The Fick's law of diffusion states that flux due to diffusion is proportional to concentration gradient,
\begin{equation}
\vec{F}_{\rm{diff}}= -\bf{K}\nabla C, \nonumber 
\end{equation}
where $\bf{K}$ is  the diffusivity matrix defined as
\begin{equation}
\bf{K} = \begin{bmatrix}   K^x & 0 & 0  \\  0 & K^y & 0 \\ 0 & 0 & K^z \end{bmatrix},   \nonumber 
\end{equation}
in terms of the eddy diffusivity coefficients $K^x, K^y, K^z$   in the $x,y$ and $z$ direction, respectively, after neglecting the molecular diffusivity.  The eddy diffusion coefficients are function of position vector.
\begin{equation}
\nabla. \vec{F}_{\rm{diff}} =  K^x C_{x,x} + K^y C_{y,y} + K^z C_{z,z}.
\label{f_diff}
\end{equation}

The second transport phenomenon is the advection which is the bulk transport of particles by a moving fluid. In our case, the driving force responsible for  particles' transportation is wind which is artificially applied.  If the flow velocity is represented by $\vec{v}=[u^x,u^y,u^z]$, we can express the advective flux as,
\begin{equation}
\label{Advection}
\vec{F}_{\rm{adv}} = \vec{v}C \nonumber 
\end{equation}
Then, the divergence of $\vec{F}_{\rm{adv}}$ is found to be,
\begin{equation}
\label{Advectionder}
\nabla . \vec{F}_{\rm{adv}} = \nabla .( \vec{v}C) = C(\nabla . \vec{v}) + \vec{v} .( \nabla C)      \nonumber 
\end{equation}
For incompressible fluids whose density stays constant \cite{fluid},
$$ 
\nabla . \vec{v}=0,
$$
Thus, the change in advection flux boils down to the following expression,
\begin{equation}
\nabla . \vec{F}_{\rm{adv}} =u^x C_x + u^y C_y + u^z C_z.
\label{f_adv}
\end{equation}

After plugging the flux terms \eqref{f_diff} and \eqref{f_adv} in \eqref{Law_conv_mass} and rearranging few terms, we obtain the following main partial differential equation (PDE),  
\begin{equation} \label{PDE_main}
C_t = S + \left( K^x C_{x,x} + K^y C_{y,y} + K^z C_{z,z} \right) - \left( u^x C_x + u^y C_y + u^z C_z  \right).   
\end{equation}

In order to derive closed-form expressions, it is necessary to define the  boundary conditions and state the basic assumptions as follows:
\begin{itemize}
	\item The downwind is in $x$ direction only i.e $\vec{v}=[u,0,0]$.
	
	\item Diffusivity coefficients are equal in all directions and are a function of downwind distance $x$ only, i.e $
	K^x = K^y = K^z \triangleq K(x) $.
	
	\item Along the downwind direction, the flux due to advection is much stronger than that due to diffusion, thus, we ignore the diffusion flux along the $x$-direction, 	$ 	K^x C_{x,x} - u C_x \approx - u C_x 	$.
	
	\item The ground $z=0$ is flat and doesn't have any topographical variations.
	
	\item The pathogen-laden droplets do not penetrate the ground, $	K(x) C_z (x,y,0) = 0$.

	\item Mass is conserved, $ 	C(x,y,\infty,t) = 0, \;\; C(x,\pm \infty,z,t) = 0 \; {\rm{ and}}  \;  C(\infty,y,z,t) = 0.
	$
\end{itemize}
By considering in the boundary conditions, PDE in \eqref{PDE_main} is simplified as 
\begin{equation} \label{simple_pde}
S =C_t+ uC_x -K(x) [C_{y,y} + C_{z,z} ], 
\end{equation}
which is more tractable allowing us to derive closed-form expressions for different sources. The detailed derivations are presented in the next section.

\section{Aerosol Channel Modeling} \label{channel}
In this section we provide a detailed analysis of aerosol channel for derive its response to both continuous and jet sources. We start with the transient response for cough and/or sneeze.

 \subsection{ Transient Response of Physical Channel to Jet Sources}

In this section, we focus on modeling of wind-aided aerosol transmission channel to understand its dynamics in both temporal and spatial domain. It can be observed that for a given location, the source for a sneeze/cough is simply an impulse. Thus, finding the response for such input will yield impulse response that can be used to compute the system performance for several input scenarios. We start our analysis by considering a single jet-source in \eqref{simple_pde} and obtain the following PDE
\begin{equation} \label{simplified_pde}
R_s \delta(x) \delta(y) \delta(z-H)\delta(t) =C_t+ uC_x -K(x) [C_{y,y} + C_{z,z}],
\end{equation}
which is written equivalently as,
\begin{align} \label{pde_to_solve}
\begin{split} 
\frac{u}{K(x)} C_x + \frac{1}{K(x)}C_t &= C_{y,y}+C_{z,z},\;\;    x\neq 0 ,\; y\neq 0 \; z\neq H  \\
C(0,y,z,t) &=\frac{ R_s}{u}  \delta(y) \delta(z-H)\delta(t).
\end{split}
\end{align}
First, we introduce the following variable definition 
\begin{equation}\label{elim}
C(x,y,z,t) = \phi (x,y,z,\tau (x,t))
\end{equation}
where, $\tau (x,t) = x-ut$. Then, we apply chain rule to find out the derivatives  $C_t = -u\phi_{\tau}$ and $C_x = \phi_{\tau}+\phi_x$ and plug them in \eqref{simple_pde} obtaining 
\begin{equation}\label{pde_solve_2}
\frac{u}{K(x)}\phi_{x} = \phi_{y,y} +\phi_{z,z}.
\end{equation}
To get rid of variable coefficient of $\phi_x$, we define,
$ x_\eta  \triangleq  \frac{u}{k(x)} $, which allows us to rewrite \eqref{pde_solve_2}  as
\begin{equation}
\label{PDE_changed}
\phi_\eta(\eta,y,z,\tau)  = \phi_{y,y}(\eta,y,z,\tau)+\phi_{z,z}(\eta,y,z,\tau).
\end{equation}
To solve \eqref{PDE_changed}, we assume $ \phi(\eta,y,z,\tau)$ is composed of separable functions as 
\begin{equation}
\label{broken_PDE}
   \phi(\eta,y,z,\tau) \triangleq   F(\eta,y)G(\eta,z)H(\eta,\tau), 
\end{equation}
which enables use to break  \eqref{PDE_changed}  into three sets of simpler PDEs with boundary conditions as, 
\begin{multicols}{3}
	\noindent
	\begin{subequations}\label{M}
		\vspace*{-1cm}
		\begin{flalign}
		F_\eta =& F_{z,z}  \label{M1}\\ 
	\label{M2}	F(0,z)  =& \frac{R_s}{u}\delta (z-H)  \\
	\label{M4}	F(\infty,z) & =0 \\
		 \label{M5} F_z(\eta,0) & =0 \\
		F(\eta,\infty) & =0  
		\end{flalign}
	\end{subequations} 
	\begin{subequations}\label{N}
		\begin{flalign}
	\label{N1}	G_\eta &= G_{y,y}  \\
		G(0,y) & = \delta (y) \\
		G(\infty,y) & =0 \\
		G(\eta,\pm \infty) & =0.  
		\end{flalign}
	\end{subequations}
\begin{subequations}\label{P}
	\begin{flalign}
	H &=\phi(\eta,\tau) \\
	H(0,t) & = \delta (t) \\
	H(\infty,t) & =0 \\
	H(\eta,\pm \infty) & =0.  
	\end{flalign}
\end{subequations}
\end{multicols}

Now we use Laplace transform to solve the above differential equations to find $F$, $G$ and $H$. At this point we assume that $F$, $G$ and $H$ are real-valued positive functions whose Laplace transform exists. The Laplace transform variables for $\eta$, $y$ and $z$ are $q$, $r$ and $s$ respectively. We start with the equation set \eqref{M} to find $F$ by finding the Laplace transform of \eqref{M1} w.r.t.~$z$, 
\begin{equation} \label{M1-A}
\bar{F}^{(z)}_\eta(\eta,s) = s^2 \bar{F}^{(z)}(\eta,s) - sF(\eta,0)-F_z(\eta,0).  
\end{equation}
Since, the droplets do not penetrate the ground, from \eqref{M5}, the last term becomes zero. Then, we take Laplace transform for  \eqref{M1-A} w.r.t. $\eta$ obtaining
\begin{equation} \label{eq3}
q\bar{\bar{F}}^{(z,\eta)}(q,s)-\bar{F}^{(z)}(0,s) = s^2 \bar{\bar{F}}^{(z,\eta)}(q,s) - s \bar{F}^{(\eta)}(q,0).
\end{equation}
By taking Laplace of \eqref{M2} we obtain $\bar{F}^{(z)}(0,s)= \frac{R_s}{u}e^{-sH}$, which is plugged in \eqref{eq3} obtaining
\begin{equation}  \label{eq3_1}
\bar{\bar{F}}^{(z,\eta)}(q,s) = \frac{ s \bar{F}^{(\eta)}(q,0)}{s^2 - q} - \frac{R_se^{-sH}}{u(s^2 - q)}.
\end{equation}
Then, by taking the inverse Laplace transform of \eqref{eq3_1} w.r.t $s$, we get,
\begin{equation} \label{eq4}
\bar{F}^{(\eta)}(q,z) = \bar{F}^{(\eta)}(q,0) \cosh(\sqrt{q}z) - \frac{R_s}{u\sqrt{q}}\sinh(\sqrt{q}(z-H)).  
\end{equation}
Now, we use the boundary condition \eqref{M4} to obtain
\begin{equation}
\bar{F}^{(\eta)}(q,0) = \frac{R_s e^{\sqrt{q}(z-H)}}{u\sqrt{q}e^{\sqrt{q}z}} = \frac{R_s}{u\sqrt{q}}e^{-\sqrt{q}H},
\nonumber
\end{equation}
which is plugged in \eqref{eq4} and after expanding the hyperbolic functions as sum of exponentials the following simplified expression is obtained,
\begin{equation}
\bar{F}^{(\eta)}(q,z) = \frac{R_s}{2u\sqrt{q}} ( e^{-\sqrt{q}(z-H)} + e^{-\sqrt{q}(z+H)} ).
\nonumber
\end{equation}
Then, we take the inverse Laplace transform w.r.t $q$ obtaining
\begin{equation} \label{fsoln}
F(\eta,z) = \frac{R_s}{u\sqrt{4\pi \eta}} \left( e^{-\frac{(z-H)^2}{4\eta}} + e^{-\frac{(z+H)^2}{4\eta}} \right).
\end{equation}
It should be noted that the domain of $z$ is above ground i.e $z>0$. Now, we consider the next set of equations \eqref{N} and take Laplace of \eqref{N1} w.r.t $y$ and $\eta$ . After using suitable boundary conditions and rearranging, the following equation is obtained
\begin{equation}
\bar{\bar{G}}^{(y,\eta)}(q,r) = \frac{ r d_1}{r^2 - q} - \frac{d_2}{r^2 - q},
\end{equation}
where $d_1 =  \bar{G}^{(\eta)}(q,0)$ and $d_2 = 1-\bar{G}_y(q,0)$. Then, by taking the inverse Laplace transform w.r.t $r$, we find
\begin{equation}\label{N_inter}
\bar{G^{\eta}}(q,y) = d_1 \cosh(\sqrt{q}y)- \frac{d_2}{\sqrt{q}} \sinh(\sqrt{q}y).
\end{equation}
After expanding the hyperbolic functions and considering  (\ref{N}d), we obtain
$$
\bar{N^{\eta}}(q,\infty) = \frac{d_1}{2} e^{\sqrt{q}y}-\frac{d_2}{2\sqrt{q}}e^{\sqrt{q}y}=0
$$
thus, $ d_1 = \frac{d_2}{\sqrt{q}}$, which can be substituted in \eqref{N_inter} obtaining
$
\bar{G^{\eta}}(q,y) = \frac{d_2}{\sqrt{q}}e^{-\sqrt{q}y}.
$
Following same strategy used for  $\bar{F}^{(\eta)}(q,z)$, we take inverse Laplace of $\bar{G^{\eta}}(q,y)$ yielding,
\begin{equation}
G(\eta,y) = \frac{d_2}{\sqrt{\pi\eta }}e^{-\frac{y^2}{4\eta}}.
\end{equation}
To find $d_2$, we use (\ref{N}b) that gives
\begin{equation}\label{d2_delta}
\delta(y) =  \lim_{\eta \to 0} \frac{d_2}{\sqrt{\pi\eta }}e^{-\frac{y^2}{4\eta}}.
\end{equation}
Then, by using the definition of delta as a limit of Gaussian,
$\delta(y) =  \lim_{x \to 0} \frac{1}{\sqrt{2\pi  x^2}}e^{-\frac{y^2}{4x^2}}
$, and comparing with \eqref{d2_delta}, we find
$
d_2 = \frac{1}{2}.
$
Therefore, $G(\eta,y)$ turns out to be,
\begin{equation}\label{Gsoln}
G(\eta,y) =   \frac{1}{2\sqrt{\pi\eta }}e^{-\frac{y^2}{4\eta}}.
\end{equation}
Now we consider the third set \eqref{P}, where the solution to our PDE is of the form $ H = \phi(\eta,x-ut)
$. Now we deploy the initial condition (\ref{P}(b)) to find the exact solution,
$$
\lim_{\eta \to 0}H(\eta,t) = \delta(t) = \lim_{\eta \to 0} \frac{1}{\sqrt{2\pi  \eta^2}}e^{-\frac{t^2}{4\eta^2}}
$$
$$
\lim_{\eta \to 0} \phi(\eta,x-ut) = \lim_{\eta \to 0} \frac{u}{\sqrt{2\pi  \eta^2}}e^{-\frac{(x-ut)^2}{4\eta^2}}
$$
Comparing both sides of the equation yield the exact solution of \eqref{P},
\begin{equation}\label{Hsoln}
H(\eta,t) = \frac{u}{2\sqrt{\pi \eta}}e^{-\frac{(x-ut)^2}{4\eta}}.
\end{equation}  
The general concentration  expression is found by plugging \eqref{fsoln}, \eqref{Gsoln}, \eqref{Hsoln} in \eqref{broken_PDE}~yielding,
\begin{equation}
C(x,y,z,t) =\frac{R_s}{8(\pi\eta)^{3/2}}e^{\frac{-(x-ut)^2}{4\eta}}e^{-\frac{y^2}{4\eta}} ( e^{-\frac{(z-H)^2}{4\eta}} + e^{-\frac{(z+H)^2}{4\eta}} ).
\end{equation} 
For ease of notation, we have just written $\eta$ instead of $\eta(x)$. However, it should be implicit that $\eta$ is a function of $x$. 
\subsection{Transient Response of Breath  and Multiple Sources}
After deriving an expression for system response to cough or sneeze, we move to breathing which is a continuous source of aerosols. The solution derived in the previous section is shown to be \textit{linear} and \textit{time-invariant} as shown in the Appendix. Thus, for extending the above analysis to evaluate system performance for different sources, we observe that for a given point in space the concentration of aerosols can be expressed as a convolution in time. Hence, we can conclude that for a  point $\vec{r}=[x,y,z]$ in space, the \textit{impulse} response of system is,
\begin{equation}\label{ht}
h_{\vec{r}}(t) = \frac{1}{8(\pi\eta)^{3/2}}e^{\frac{-(x-ut)^2}{4\eta}}e^{-\frac{y^2}{4\eta}} ( e^{-\frac{(z-H)^2}{4\eta}} + e^{-\frac{(z+H)^2}{4\eta}} )
\end{equation}
As explained in section \ref{input}, breathing can be modeled as a unit step function of magnitude $R_b$. The output for this breathing source can be obtained by convolving with the impulse response,
$$
y_{\vec{r}}^{\mathrm{b}}(t) =\int_{-\infty}^t \alpha e^{\frac{(x-u(t-\tau))^2}{4\eta}} u(\tau)d\tau
$$
where $\alpha= \frac{1}{8(\pi\eta)^{3/2}}e^{-\frac{y^2}{4\eta}} ( e^{-\frac{(z-H)^2}{4\eta}} + e^{-\frac{(z+H)^2}{4\eta}} )$

The final response can be expressed in terms of complementary error function,
\begin{equation}\label{br}
y_{\vec{r}}^{\mathrm{b}}(t) = \frac{R_{\mathrm{b}}}{8\pi\eta u } \bigg\{ \mathrm{erfc}(\frac{x-ut}{2\sqrt{\eta}}) - \mathrm{erfc}(\frac{x}{2\sqrt{\eta}}) \bigg\} e^{-\frac{y^2}{4\eta}} \big( e^{-\frac{(z-H)^2}{4\eta}} + e^{-\frac{(z+H)^2}{4\eta}} \big).
\end{equation}
As mentioned in Section \ref{input}, for a given person the source is an additive function of continuous and jet sources. Since, the system is linear, the overall response of the system for a person of height $H$ who is standing at origin and is coughing or sneezing at some random instants $t_i$ is given by,
\begin{equation}
y_{\vec{r}}(t) \!\! = \!\! \Bigg[ \frac{R_{\mathrm{b}}}{8\pi\eta u } \bigg\{ \mathrm{erfc} (\frac{x-ut}{2\sqrt{\eta}}) -\mathrm{erfc} (\frac{x}{2\sqrt{\eta}}) \bigg\}  
  + \sum_{i=1}^{N} \frac{R_{\mathrm{s}}}{8(\pi\eta)^{3/2}}e^{\frac{(x-u(t-t_i))^2}{4\eta}} \Bigg]e^{-\frac{y^2}{4\eta}} \big( e^{-\frac{(z-H)^2}{4\eta}} + e^{-\frac{(z+H)^2}{4\eta}} \big)
\end{equation}
where $N$ is the number of jet sources, i.e coughs and sneezes. 

In a similar way, the response of multiple user scenario is found to be 
\begin{equation}\label{multi}
y_\mathrm{multi}(t) = \sum_{j=1}^{N_{\mathrm{u}}} y_{\vec{r_j}}^{t_j}(t)u(x-x_j)u(t-t_j)
\end{equation}
where $\vec{r}_j=[x-x_j,y-y_j,z-z_j]$ and $N_{\mathrm{u}}$ is the number of users, $x_j$, $y_j$ and $z_j$  are coordinates of the $j-\mathrm{th}$ user's source and $t_j$ is the releasing time. A stochastic setting can also be developed by discretizing the time   equally with an interval $T_1$ and defining the probability that a user $j$ sneezes in the interval $T_i$ by $p_{ij}$. It can assumed that the intervals are quiet small and therefore for each interval the sneeze/cough time can be taken to be start of interval i.e $T_i$. In this case the system response at time $T$ is given by 
\begin{equation}
y_\mathrm{multi}(t) = \sum_{i=1}^{L} \sum_{j=1}^{K} p_{ij} y_{\vec{r_j}}^{t_j}(t)u(x-x_j)u(t-t_j),
\end{equation}
where $ L= \left \lceil{\frac{T}{T_1}}\right \rceil$.
\subsection{Steady State Response of Breath}
To derive the  steady state response of emitted pathogens from a continuous source, i.e. breathing, we start from   \eqref{simple_pde} and consider the following changes:
\begin{enumerate}	
	\item The infected human is continuously emitting pathogens at a constant rate $R$ (discussed in section III.A). The person is standing at origin and his mouth is at height $H$. Thus, we can represent the source concentration as,
	\begin{equation}
	S\left(\vec{r},t \right) = R \delta(x) \delta(y) \delta(z-H).   \nonumber
	\end{equation}
	\item The solution is derived for steady state conditions therefore $C_t =0$.
\end{enumerate}
After considering these conditions, \eqref{simple_pde}  reduces to
\begin{equation} \label{PDE_simplified}
	R \delta(x) \delta(y) \delta(z-H) + K(x) [C_{y,y} + C_{z,z} ] = uC_x .
\end{equation}
Similar to the transient analysis scenario, we rewrite  \eqref{PDE_simplified} as,
\begin{equation} \label{PDE_eqv}
\begin{split}
\frac{u}{K(x)} C_x = C_{y,y}+C_{z,z} \\
uC(0,y,z) = R \delta(y)\delta(z-H),
\end{split}
\end{equation}
which is a variable coefficient PDE. To simplify \eqref{PDE_eqv}, we adopt the same procedure as followed earlier and define a new variable  $\eta$, where $ \eta = \frac{1}{u}\int_0^x K(t) dt$. Thus,  the PDE in \eqref{PDE_eqv} reduces to
\begin{equation} \label{simp_PDE}
C_\eta = C_{y,y}+C_{z,z}.   
\end{equation} 
Then, we assume $C(\eta,y,z) $ is separable function as
\begin{equation} \label{PDE_break}
C(\eta,y,z) =  F(\eta,y) G(\eta,z),
\end{equation}
which allow breaking the PDE in \eqref{simp_PDE} into two sets of simpler PDEs as follows 
\begin{multicols}{2}
	\begin{subequations}
		\begin{align}
		F_\eta &= F_{z,z} \label{eq1a} \\
		F(0,z) & = \frac{R}{u}\delta (z-H) \label{eq1b} \\
		F(\infty,z) & =0  \label{eq1c} \\
		F(\eta,\infty) & =0 \label{eq1d}  \\
		F_z(\eta,0) & =0 \label{eq1e}.
		\end{align}
	\end{subequations}
	\begin{subequations}
		\begin{align} 
		\nonumber \\
		G_\eta &=G_{y,y}  \label{eq2a} \\
		G(0,y) & = \delta (y) \label{eq2b}  \\
		G(\infty,y) & =0 \label{eq2c} \\
		G(\eta,\pm \infty) & =0.  \label{eq2d}  		
		\end{align}
	\end{subequations}
\end{multicols}
Since the set of simplified PDEs and boundary conditions is the same as \eqref{M} and \eqref{N}, it is pretty straightforward to see that the solution to \eqref{simp_PDE} is given by,
\begin{equation} \label{C_eqn}
C(\eta,y,z)=   \frac{R}{4u\pi \eta} e^{-\frac{y^2}{4\eta}} \left( e^{-\frac{(z-H)^2}{4\eta}} + e^{-\frac{(z+H)^2}{4\eta}} \right),
\end{equation}
 which resembles a Gaussian curve with a variance parameter of $ 2\eta$ that describes the spread of the concentration curve in crosswind directions and is an indicator of turbulence present in the environment due to the wind.

\begin{figure}[t!]
	\centering
	\includegraphics[width=4.5in]{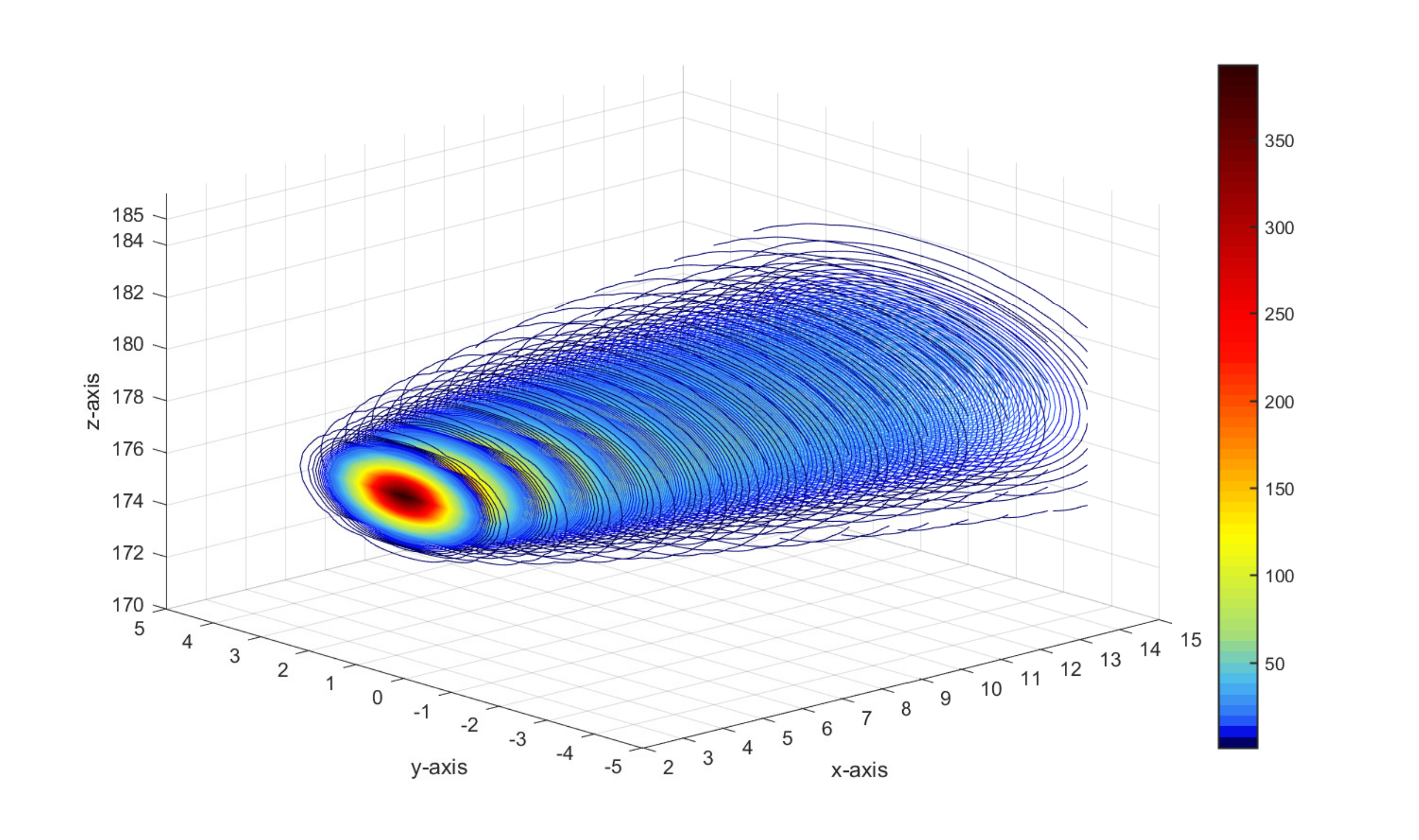}
	\caption{Concentration of virus for source present at $\vec{r}=[0,0,177]$}
	\label{con_v}
\end{figure}
 
\begin{figure}[t!]
	\centering
	\includegraphics[width=4.5in]{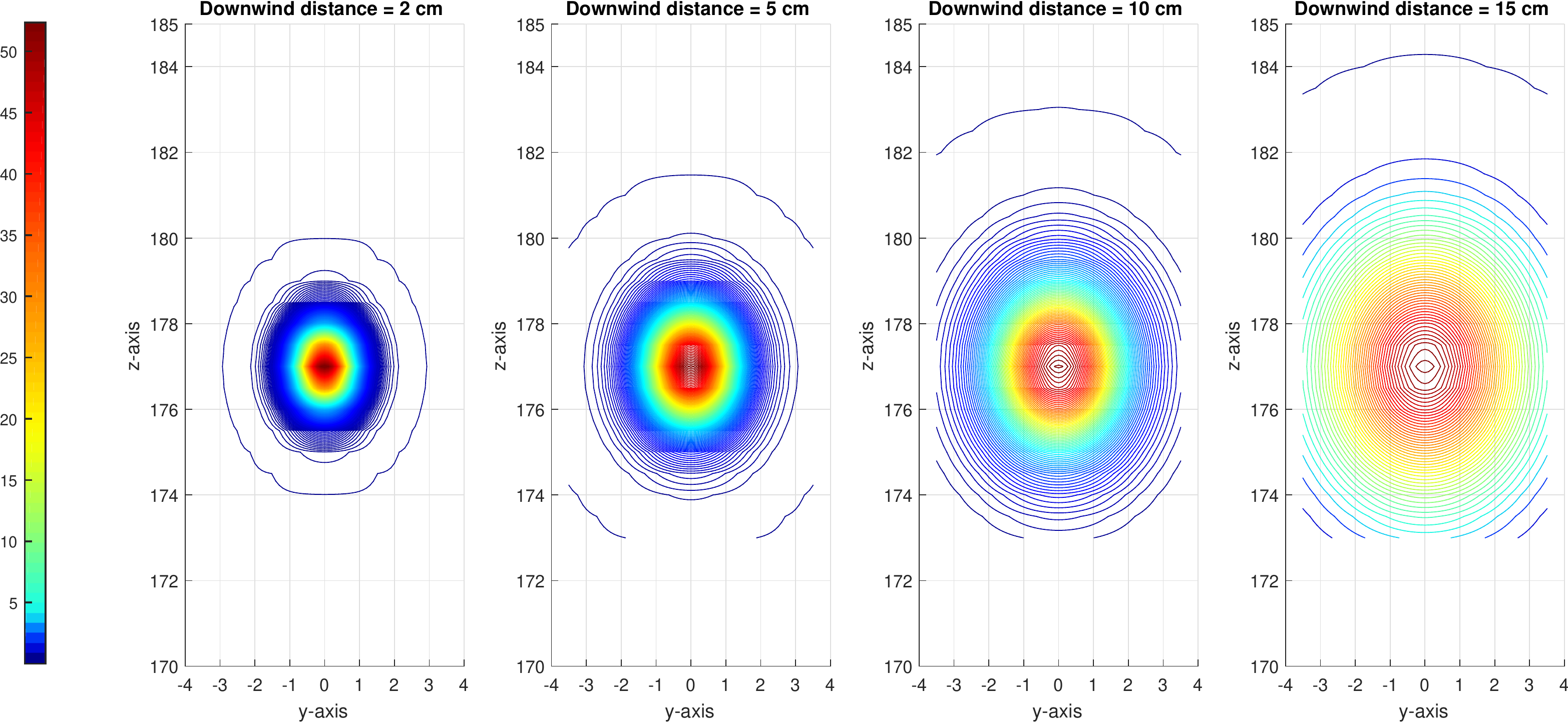}
	\caption{Impact of downwind distance on virus concentration}
	\label{dis_v}
\end{figure}

To visualize the spread of aerosols, a contour plot is presented in Fig. \ref{con_v}. The source is located at a certain height at origin which can be seen as the point of highest concentration. Moving away from the source results in a drop in concentration, however, this change is not constant. In fact, it is like a cone with it's vertex at the source spreading in the direction of wind. Analyzing the expression derived above, it can be observed that the the expression is in fact product of two normal curves whose variance changes with downwind distance. This trend can be more clearly visualized in Fig. \ref{dis_v} where increase in downwind distance increases the spread of individual contours in the $y-z$ plane. Also this spread is more significant up till a certain point along the x-axis after which the overall or mean concentration itself is too low to be sensed or detected.   Thus, it can also be concluded that on average the concentration drops as we move away from the source, however the trend is not linear. 

\subsection{Frequency Response}

In Section IV-B, we have focused on the temporal behavior of channel and therefore observing the frequency response of system can provide useful insights into the behavior of channel. Also, since the objective of this work is to view this aerosol dispersion phenomenon in the context of a communication systems problem, it is worthwhile to analyze the aerosol channel's frequency response. Furthermore, in order to deal with the issue of synchronization at receiver end and choose the appropriate sampling frequency, it is necessary to study the system characteristics in frequency domain. 

The Fourier transform of the system  impulse response  given by \eqref{ht} can be found  from  
$$
H(\omega) = \alpha \int_{-\infty}^{\infty} e^{-\frac{(x-ut)^2}{4\eta}}e^{-j\omega t}dt$$ 
where $\alpha = \frac{1}{8(\pi\eta)^{3/2}}e^{-\frac{y^2}{4\eta}} ( e^{-\frac{(z-H)^2}{4\eta}} + e^{-\frac{(z+H)^2}{4\eta}} )$. After completing the square and rearranging the exponent terms, we obtain the following expression,
$$
H(\omega) = \alpha e^{-(\frac{\omega^2\eta}{u^2} + \frac{j\omega x}{u}  )} \int_0^{\infty} e^{-(\frac{ut}{2\sqrt{\eta}} - \frac{x}{2\sqrt{\eta}} -\frac{j\omega \sqrt{\eta}}{u} )^2 }dt.
$$
Then, we apply a change of variables and obtain the following simplified expression,
\begin{equation}\label{hf}
H(\omega) = \frac{f(y,z)}{8u\eta\sqrt{\pi}} e^{-(\frac{\omega^2\eta}{u^2} + \frac{j\omega x}{u}  )},
\end{equation}
where $f(y,z)=e^{-\frac{y^2}{4\eta}} ( e^{-\frac{(z-H)^2}{4\eta}} + e^{-\frac{(z+H)^2}{4\eta}} )$.

\begin{figure}
	\begin{subfigure}{.5\textwidth}
		\includegraphics[scale=0.45]{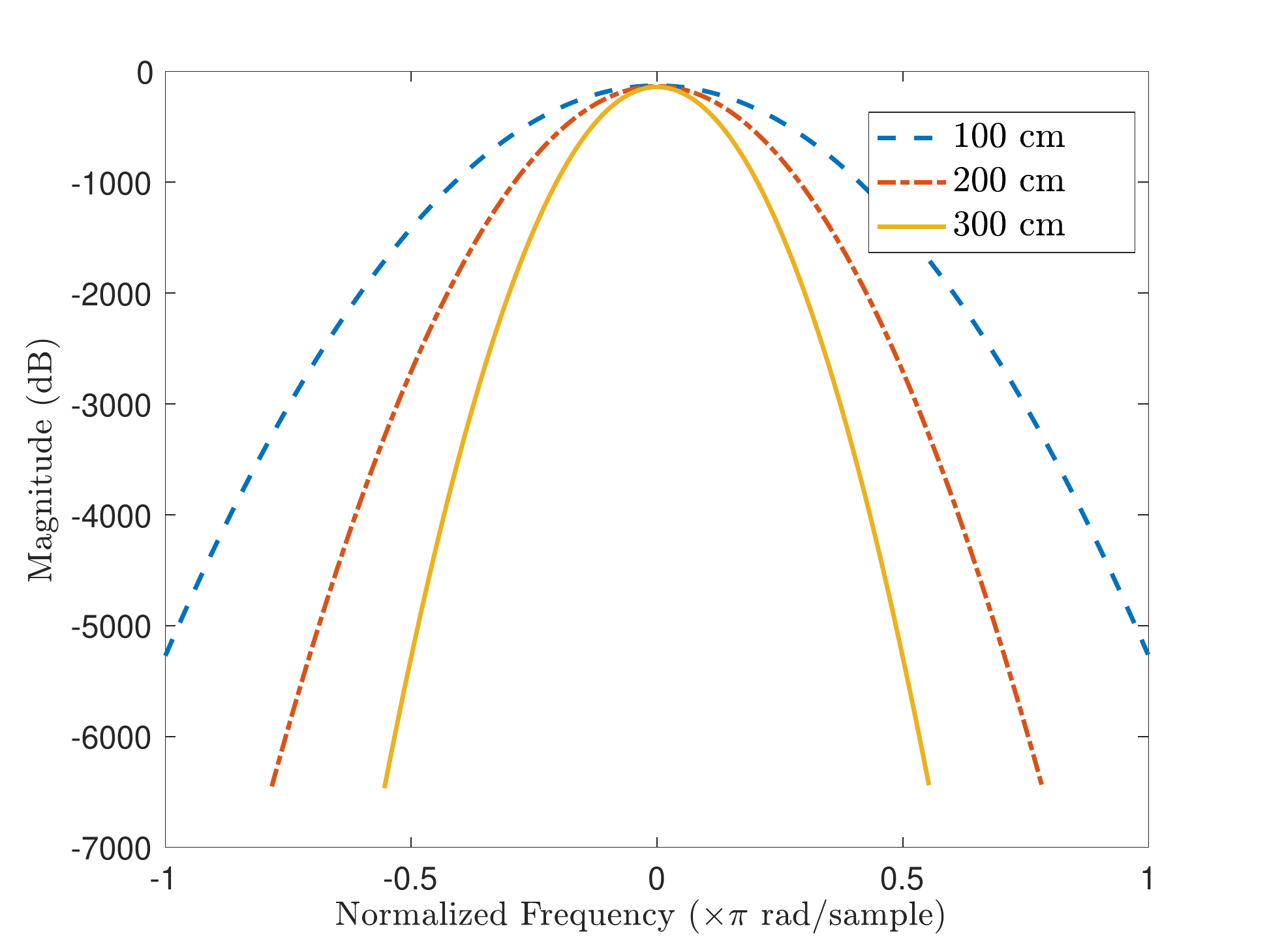}
		\caption{a. Magnitude Response for different downwind distance}
		\label{fig:f1}
	\end{subfigure}
	\begin{subfigure}{.5\textwidth}
		\includegraphics[scale=0.45]{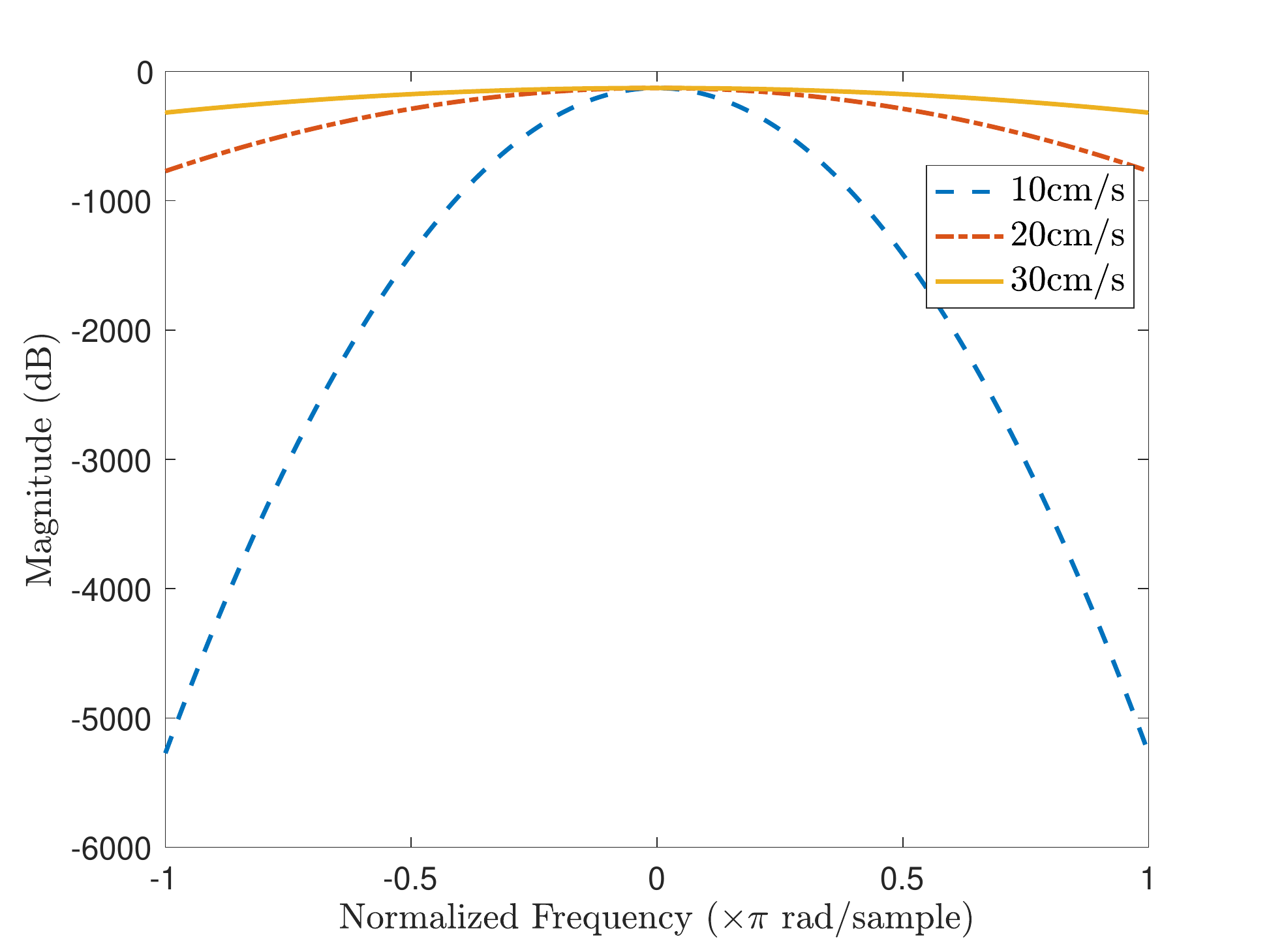}
		\caption{b. Magnitude Response  for different wind speeds}
		\label{fig:f2}
	\end{subfigure}
	\caption{Frequency response of aerosol channel}
	\label{fig:fre}
\end{figure}
We can visualize the impact of distance and wind speed on frequency response  in Fig \ref{fig:f1} and \ref{fig:f2}. It can be seen that the channel becomes more selective at lower wind speed and/or large observation distances.

\section{Detection Performance Analysis}
The viral aerosol detection system performance is affected by several factors such as, the diffusion process,  Brownian motion, interference and the electrical properties of receptors.  In order to analyze the effect of diffusion in-depth, stochastic models for the aerosol channel need to be derived, which is a complicated and extensive process because of the complexity associated with the modeling of turbulent fluid flows. A simpler approach at this point is to assume a single viral source to ignore the interference effect, then evaluate the average response, which is what we derived in section III-B. For the electrical properties of receptors or Si-NW FET in particular, there is sufficient literature on noise modeling that depends on electrical properties of FETs \cite{nse1}. However, since we want to focus on the spread of aerosols and their integration in a communication system, instead of diving deep into a well-detailed model particular for Si-NW FET, we opt for the basic model and as mentioned previously use an additive Gaussian noise. It must be noted that approximating the non-signal dependent noise term to be Gaussian is a good assumption. Since, the receiver consisting of thousands of receptors adds the input, that has finite variance measurement noise,  from each receptor separately, this independence allows us to apply central limit theorem. Hence, the final expression for the average received concentration incorporating the noise effects can be written~as,
\begin{equation}
\label{final_eq}
C_{\rm{r}}  = \xi  \gamma C_{\rm{mean}} + n 
\end{equation} 
where $\xi$ represents aerosol sampler efficiency, $\gamma$ is a fraction representing the probabilistic binding process, $n$ is additive  noise that incorporates the effect of flicker, thermal noise and other residual noise and is modeled as Gaussian \cite{sinw2} with mean zero and variance $\sigma^2$.   Although the exact expression for $\sigma^2$ can be derived, it is a tedious process. In order to observe the system performance and trends, it is sufficient to treat the noise variance as a parameter that can be varied over a range. Moreover, the advancements in nanotechnology and electronics provide us  sufficient flexibility in design of MC receivers. Compared to the transmitter and physical channel where there is little or almost no flexibility in design process, there is a lot of room in the physical design of molecular receiver.  $C_{\rm{mean}}$ is average concentration of virus particles that are present in the receiver chamber and is given by,
\begin{equation} \label{C_mean}
C_{\rm{mean}} = \int_{T_{\mathrm{s}}}  \int \int \int_{V_{\rm{rx}}} C\left( x,y,z,t \right)  dx dy dz dt
\end{equation} 
where $V_{\rm{rx}}$ represents receiver volume and $T_{\mathrm{s}} $ is the sampling time period $T_{\mathrm{s}} $ is very large, \eqref{C_mean} reduces to the one with steady state performance.    As for the receive, it receiver could be
a 1-D array of biosensors, a sphere or some other appropriate surface molecular receptors for corresponding viral aerosols. It is worthy noting  that the receptor's size and density determine the hitting probability which eventually change the binding noise. It has been shown for diffusion-based MC that higher number of smaller receptors improve the hitting probability \cite{receptor}. In our system, the effect of receptor size and density is implicitly captured through $\gamma$, however, the impact of the receptor density, size and geometry can be analyzed in detail in future studies.  Depending upon the receiver geometry and receptor/biosensor density,
closed form expressions for average received concentration
of pathogen may also be derived. However, the process is quiet complex and closed-form expressions might not exist in all scenarios for which data-driven approaches and machine learning tools can be employed.

In the detection process, $C_{\rm{mean}}$ is compared with a pre-determined threshold, $C_{\rm{th}}$,  through maximum posterior probability rule to determine  whether the person is infected or not. Let $\mathcal{I}$ and $\mathcal{H}$ represent the infected and healthy person, respectively. Thus,  the decision is made according~to,

\begin{equation}
\Pr(\mathcal{I} |C_{\rm{r}}) \lessgtr^{^\mathcal{H}}_{_\mathcal{I}} \Pr(\mathcal{H}|C_{\rm{r}}).  \nonumber
\end{equation}
Then, by applying Bayes' rule and assuming that the event of person being infected or healthy is equally likely, we obtain the following expression,
\begin{equation}
\Pr\left(C_{\rm{r}}|{\mathcal{I}}\right) \lessgtr^{^\mathcal{H}}_{_\mathcal{I}} \Pr \left(C_{\rm{r}}| \mathcal{H} \right),  \nonumber
\end{equation}
 which can be rewritten equivalently using \eqref{final_eq}  as,
\begin{equation}
\frac{1}{\sqrt{2\pi  \sigma^2 } } e^{\frac{-(C_{\rm{r}}- \gamma \xi C_{\rm{mean}} )^2}{2  \sigma^2}} \lessgtr^{^\mathcal{H}}_{_\mathcal{I}} \frac{1}{\sqrt{2\pi  \sigma^2} } e^{\frac{-(C_{\rm{r}} )^2}{2 \sigma^2}}.  \nonumber
\end{equation}
After rearranging the terms on both sides, we obtain the following inequality,
\begin{equation}
C_{\rm{r}} \lessgtr^{^\mathcal{H}}_{_\mathcal{I}}\frac{  \gamma \xi C_{\rm{mean}}}{2}.
\end{equation}
Thus, the maximum likelihood threshold value should be equal to,
\begin{equation} \label{Cth}
C_{\rm{th}} =\frac{  \gamma \xi C_{\rm{mean}}}{2}.
\end{equation}
For further insights, we derive the probability of missed detection $P_{\rm{md}}$ as follows,
\begin{equation}
\label{Pm}
P_{\rm{md}} = \Pr \left(C_{\rm{r}} \leq C_{\rm{th}}|\mathcal{I} \right),
\end{equation}
which reduces  to  
$$
P_{\rm{md}} = \int_{-\infty}^{C_{\rm{th}}} \frac{1}{\sqrt{2\pi}N_o} e^{\frac{-(C_{\rm{r}}- \gamma \xi C_{\rm{mean}} )^2}{2 \sigma^2}}.
$$
Through change of variables, we get the final expression for probability of missed detection in terms of the well-known $Q$-function,
\begin{equation}
\label{Pmf} 
P_{\rm{md}} = Q \left(\frac{\gamma \xi {C_{\rm{mean}}}}{\sqrt{2^3  \sigma^2}} \right)
\end{equation}
where,  $Q(x) = \frac{1}{2\pi} \int_{x}^{\infty} e^{-\frac{u^2}{2}}du$.

From \eqref{Pmf} and \eqref{Cth}, it can be observed that the probability of missed detection not only depends on the average received a concentration of pathogens but also on the sampler efficiency on which is taken into account while incorporating the threshold value. 
\section{Numerical Results}
\label{result}
In this section, we analyze the  proposed system performance numerically by studying the spatial temporal viral concentration in addition the missed detection probability. Throughout the following numerical results, we assume that the receiver is a sphere of radius $r_d$ placed in-line with the source along downwind direction in the $y-z$ plane at a distance of $d_{x}$ and  centered at $\vec{u_s} = [d_{x},0,H]$ as assumed in Fig.~\ref{con_v}. Additionally, we assume the receiver is located in a perfectly sanitized environment. In the following numerical results, we adopt the parameters listed in Table \ref{tab:table1}, unless otherwise specified.
  
 \begin{table}[t!]
 	\centering
 	\caption{Simulation Parameters }
 	\label{tab:table1}
 	\begin{tabular}{l|l}
 		\toprule
 		Parameters & Values \\
 		\midrule
 		$u $ & $140 \; \mathrm{cm/sec}$  \\ 
 		
 		$H$ & $180 \; \mathrm{cm}$ \\
 		
 		$K$ & $  0.242 \;\mathrm{cm^2/sec}$ \\
 		
 		$r_\mathrm{d}$ & $2 \; \mathrm{cm}$   \\
 		
 		\bottomrule
 	\end{tabular}
 \end{table}
  
\begin{figure}[t!]
	\centering
	\includegraphics[width=3.5in]{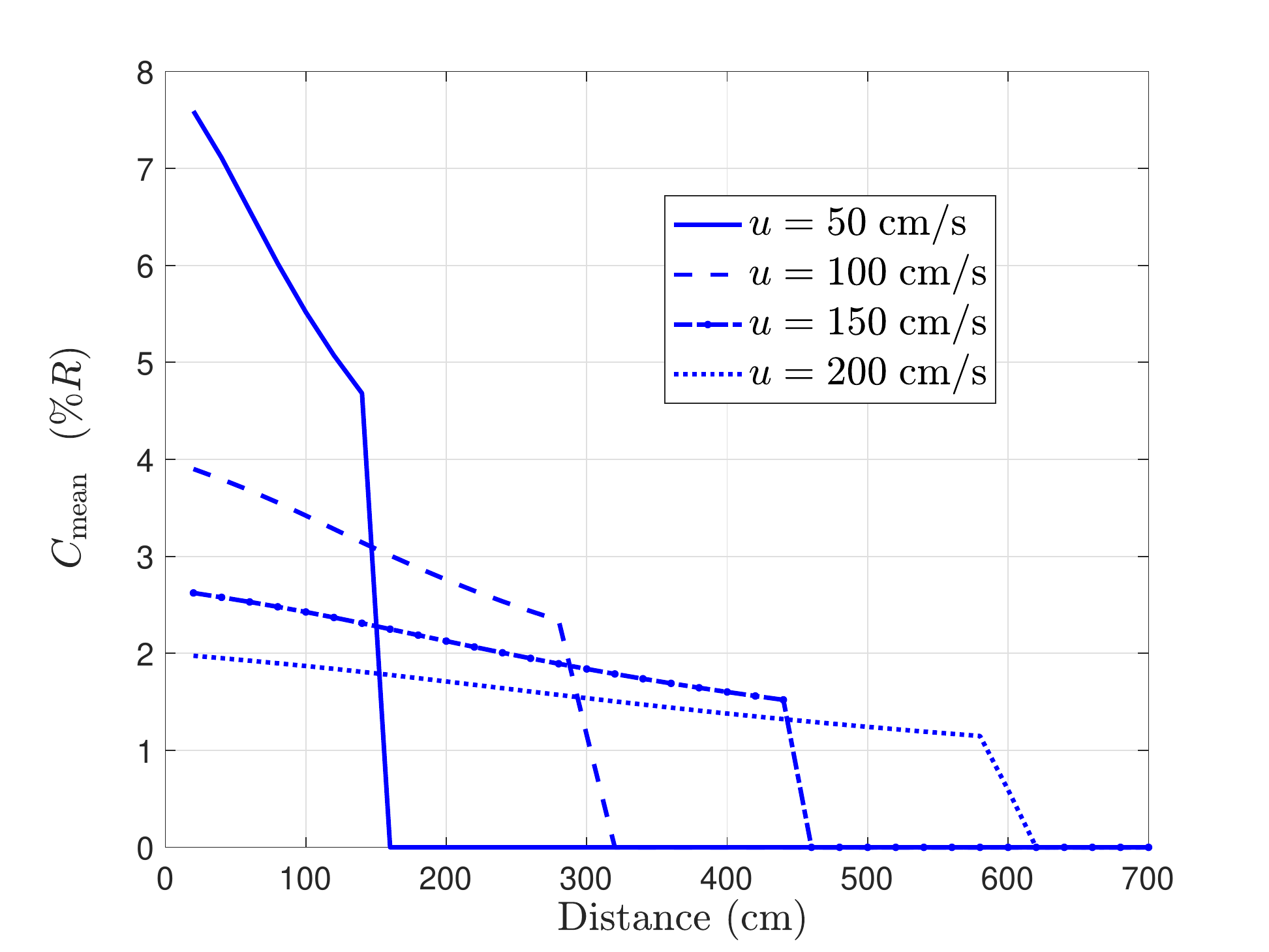}
	\caption{Detected viral concentration ratio versus the distance  between the infected person and the detector for different airflow velocities.}    
	\label{Sim_Ex1}
\end{figure}

 In the first numerical example, we study the effect of airflow velocity and distance on the received viral concentration at the receiver side. For this purpose, we assume the sampling time is $3$ seconds and evaluate the collected viral concentration performance as a ratio of the released virus $R$ versus the distance between infected person and receiver for different $u$ ranges as shown in Fig. \ref{Sim_Ex1}.  The air flow velocities are chosen to represent the scenario of exhaled nasal breath where it is below 140 cm/s, cough scenario or breath with artificial air flow \cite{tang2013airflow}. First, we observe that the spatial viral signature depends mainly on airflow velocity. The detection area can be increased by increasing $u$, however it comes at a cost of concentration decrease as a result of the mass conservation law \cite{arya1999air}. Thus, applying air flow plays a good role in extending the spatial coverage of detection, but it may reduce the concentration below detectable level.  
 
\begin{figure}[t!]
	\centering
	\includegraphics[width=3.5in]{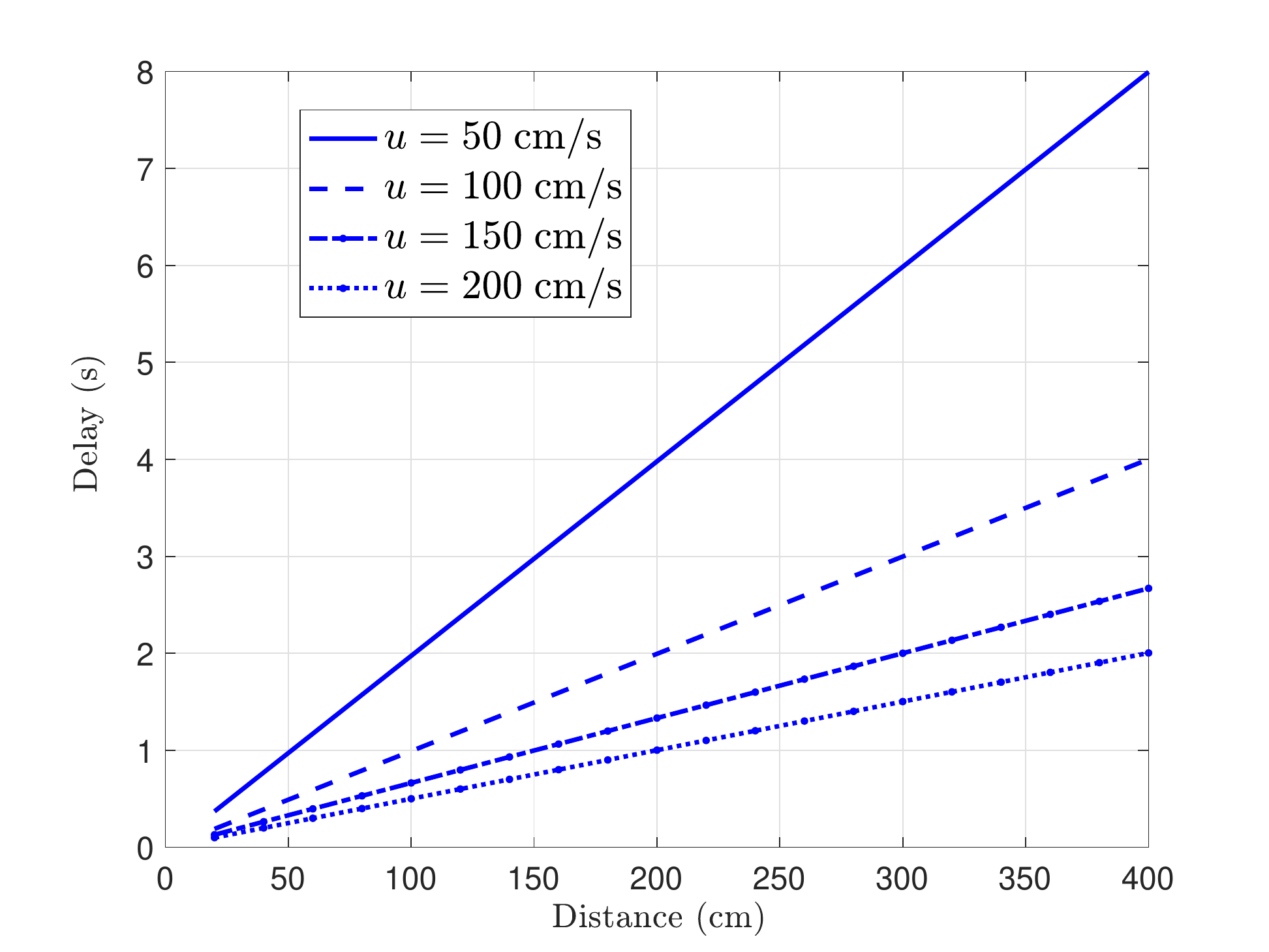}
	\caption{Propagation delay in detecting $1\%$ of  the exhaled virus versus a wide range of distances between the human and  detector for different airflow velocities. }    
	\label{Sim_Ex2}
\end{figure}

In the second numerical example, we investigate the delay due to molecular channel propagation using both  diffusion and advection mechanisms; in order to detect a virus at the detector side. To this end, we plot the time needed to detect $1 \%$ of the exhaled virus versus the propagation distance for different airflow velocities in Fig. \ref{Sim_Ex2}.  One can observe that the airflow velocity has a significant effect in reducing the propagation delay, where doubling the airflow velocity can reduce the propagation delay to the half. However, it is important to know that using airflow reduces the concentration peaks, which can affect the system reliability in detecting virsus even if the detector is located close to the infected human as shown in the first numerical results depicted in Fig. \ref{Sim_Ex1}.

\begin{figure}[t!]
	\centering
	\includegraphics[width=3.5in]{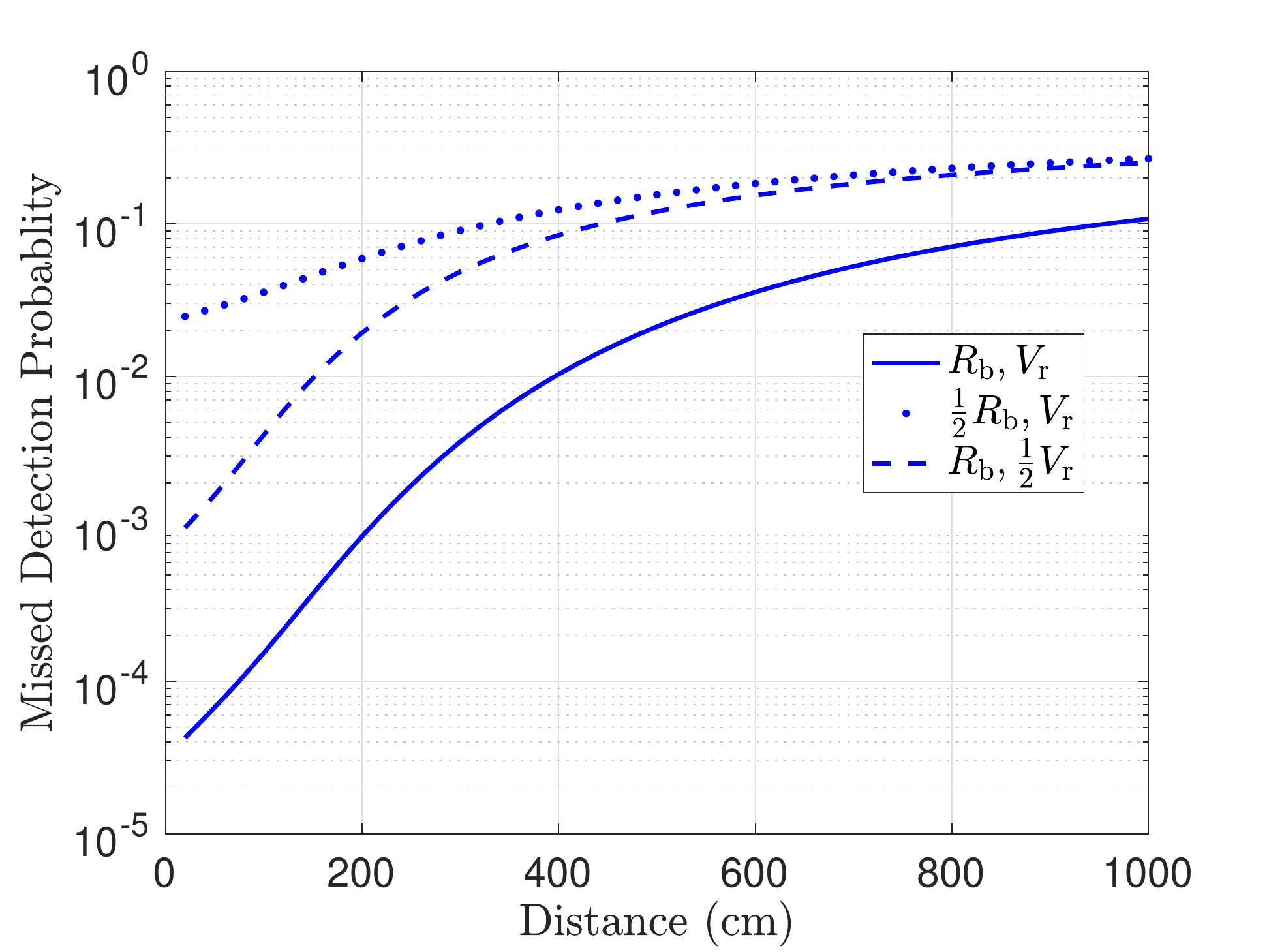}
	\caption{Effect of receiver volume and emitted virus rate on the missed detection probability versus the distance between the infected human and the detector.}    
	\label{Sim_Ex3}
\end{figure}

Finally, we study the steady state missed detection probability of exhaled nasal breath versus the distance between infected human and the detector. We assume $\frac{\xi  \gamma  R_{\mathrm{b}}}{8\sigma ^2} = 1.96\times 10^4$ and compare the missed detection probability between three scenarios: $R_{\mathrm{b}}$ virus flow rate and spherical detector with volume $V_{\mathrm{r}}$,  $\frac{1}{2}R_{\mathrm{b}}$ virus flow rate and spherical detector with volume $V_{\mathrm{r}}$ and $R_{\mathrm{b}}$ virus flow rate and spherical detector with volume $\frac{1}{2}V_{\mathrm{r}}$, as shown in Fig. \ref{Sim_Ex3}. The depicted results shows that the exhaled virus concentration ratio, detection distance and detector volume are essential parameters in determining the missed detection probability performance. Moreover, the receiver volume has better impact on the performance than virus concentration in the exhaled breath. Thus, it is desirable to design a detector with large volume(s) and appropriate air collecting capabilities. 

\section{Conclusion}
\label{conc}
The paper defined a new research dimension in MC, which is viral aerosol transmission, spread mechanism and detection. The mathematical modeling of aerosol channel provides insights into the dynamics of virus spread that can prove helpful in its detection. Applying artificial airflow overcomes the slowness of diffusion based spread and allows the detection of viruses, thus it is a key enabler for the viral detection system. The simulation results show that the missed detection is controlled by the distance, virus flow rate, air velocity, receiver binding efficiency and others. The proposed mathematical problem was studied using steady state analysis of virus transmission and detection due to breathing. It was further extended to transient analysis and response of system to coughs and sneezes. The transient analysis is important in multiple aspects of receiver design such as the memory channel behavior and synchronization. In future works, the work can be extended by relaxing the assumptions and incorporating complex wind fields, that results in different turbulence behavior. Moreover, it is imperative to optimize the receivers' size and/or location, and study multiple sources, interference, and different turbulence models. Finally, this work can also be extended in the context of predicting the occurrences of pandemics and taking preparatory measures.

\appendix

 \section{First Appendix}
\label{FirstAppendix}
In this section, we show that the system is linear and time-invariant.
\begin{itemize}
	\item \textit{Linearity}: Consider the main PDE that defines the system, 
	\begin{equation}\label{PDE}
	C_t+ uC_x -K(x)C_{y,y} + K(x)C_{z,z} =\frac{R}{u}\delta(x)\delta(y)\delta(z-H)\delta(t)
	\end{equation}
	
	It is straightforward to observe that the PDE and it's boundary conditions are linear. For linear Differential equations, the principle of superposition states\cite{PDEbook},\\
	\textit{``If a linear PDE in $c$, $L(c)=f_1$ has the solution $c_1$ and $L(c)=f_2$ has the solution $c_2$, then the solution to $L(c)=af_1+bf_2$ is given by $c_3=ac_1+bc_2$ "}
	
	Applying the principle of superposition we can conclude that for scaled and additive inputs, the output is also scaled and sum of respective responses.
	
	\item \textit{Time Invariance}
	Consider the PDE \eqref{PDE} where input is a shifted impulse in time,
	$$
	C_t+ uC_x -K(x)C_{y,y} + K(x)C_{z,z} =\frac{R}{u}\delta(x)\delta(y)\delta(z-H)\delta(t-t_o)
	$$
	Let us make a change of variable and define $t'=t-t_o$ so that the above PDE becomes,
	\begin{equation}\label{mpde}
	C_{t'}+ uC_x -K(x)C_{y,y} + K(x)C_{z,z} =\frac{R}{u}\delta(x)\delta(y)\delta(z-H)\delta(t')
	\end{equation}
	Along with the boundary condition, $C(x,y,z,t')=0\;\;\; for \;\; t'=0$.\\
	If $C_1(x,y,z,t)u(t)$ represents the solution to \eqref{PDE} then it is straightforward to observe that the solution to \eqref{mpde} is given by $C_1(x,y,z,t')u(t')=C_1(x,y,z,t-t_o)u(t_o)$. Thus, the system is time-invariant.
\end{itemize}


\end{document}